# Congestion Control With Explicit Rate Indication

Anna Charny
Digital Equipment
Corporation
charny@nacto.lkg.dec.com

David D. Clark
Massachusetts Institute
of Technology
ddc@lcs.mit.edu

Raj Jain
Ohio State University
jain@cis.ohio-state.edu

## Abstract

As the speed and the dynamic range of computer networks evolve, the issue of efficient traffic management becomes increasingly important. This work describes an approach to traffic management using explicit rate information provided to the source by the network. We present an asynchronous distributed algorithm for optimal rate calculation across the network, where optimality is understood in the maxmin sense. The algorithm quickly converges to the optimal rates and is shown to be well-behaved in transience.

## 1  Introduction

### 1.1  Background

In the past decade several mechanisms for congestion control have been developed and implemented. DECbit [34] and Slow Start [20] are perhaps the best known. Both of these schemes were developed for connectionless networks with window flow control, in which the routers had no knowledge of the individual flows and their demands, the routes changed frequently and the header space was scarce.

With the rapid increase of the ratio of propagation delay to transmission time in the modern networks, window-based schemes face significant challenges [23], [32], [33]. The large window size and long feedback delays cause large bursts of traffic to be injected into the network leading to severe congestion. As a result it has been argued that a rate-based approach may be a viable alternative to windows. A number of rate-based schemes were developed based on the basic idea of DECbit - using a single bit in the packet (cell) header, which is set by the switch in the case of congestion [36]. The sources adjust their rate up or down depending on the value of this bit received in feedback. Rate-based variations of DECnet have been developed for frame relay and fast packet networks [1], [39]. The Explicit Forward Congestion Notification (EFCN) schemes considered for ATM networks are also based on concepts originating from the DECbit scheme work.

However, the simplicity of the DECbit scheme has its price - slow convergence and rate oscillations, which can cause large queues and potential data loss in rate-based networks. Even a small overload can cause extremely high queues if sustained for a long time. Since the rate is only adjusted with a period proportional to the end-to-end round-trip delay, long-distance networks can no longer afford too many round-trips to determine correct rate allocation. Thus, a scheme with faster convergence properties is extremely desirable.

In its simplest form DECbit suffers from unfairness to individual flows. It has been long argued that some sort of selective feedback based on individual flow information at the switches is required to provide fairness [37]. Selective Binary Feedback scheme [35] is an enhancement of DECbit in which the switches calculate a fair rate allocation and set the congestion bit for only those flows which exceed the fair allocation.

This paper suggests an alternative approach to rate control, in which the switches calculate a rate allocation for the flows, and this allocation is explicitly provided to the sources in control packets or data packet headers. This approach provides significantly reduced convergence time and remarkable robustness. We identify a class of possible switch policies that lead to convergence to the optimal rate allocation. We prove an upper bound to the time of convergence, and show that the scheme is self-stabilizing, in that it will converge from any set of initial conditions. We also show that the expected performance of the algorithm during convergence is well behaved in practice.

The scheme described here is primarily suited for connection-oriented networks, in which the switches have information about the individual flows and have stable flow routes. However, it is also applicable in connectionless networks, provided that the routes do not change too often and that the individual flow information is made available to the switch.

This paper is based on the MS Thesis done by Anna Charny at MIT under the supervision of David Clark and Raj Jain [3]. Several congestion control schemes currently under consideration by the ATM Forum are based on the results of this work.

## 1.2 Network Model

We model a flow as a uni-directional data transfer from source to destination with feedback traversing the network in the opposite direction through the same route. We assume that there is a one-to-one correspondence between sources and destinations and that the routes of all flows are fixed. Flows are considered to be independent of each other. The set of flows is allowed to change dynamically.

The only assumption we make about the service discipline employed by the switch is that the packets of *each flow* are served in FIFO order. Thus, the switches could be strict FIFO, FIFO+, Priority, Stop-and-Go, Fair-Queuing, etc. ([4], [13], [33], [28], [29]).

Finally, we assume that the link capacity is the bottleneck resource in the network. However, we believe that the results of this work are equally applicable if host or switch processing capacity is the bottleneck.

## 1.3 Optimality Criterion

This work chooses the so-called *maxmin* or *bottleneck* fairness discussed in various modifications in [2], [14], [19], [31], [35].

This approach is based on the following intuition.

Consider a network with given link capacities, the set of flows, and fixed flow routes. We are interested in those rate allocations that are feasible in the sense that the total throughput of all flows traversing any link does not exceed the link's capacity. We would like the feasible rate allocation to be fair to all flows. On the other hand, we want the network to be utilized as much as possible.

We now define a fair allocation as follows. We consider all "bottleneck" links, i.e. the links with the smallest capacity available per flow. We share the capacity of these links equally between all flows traversing them. Then we remove these flows from the network and reduce all link capacities by the bandwidth consumed by the removed flows. We now identify the "next level" bottleneck links of the reduced network and repeat the procedure. We thus continue until all flows are assigned their rates.

Such a rate vector is known as a *maxmin fair* allocation. It can be seen that a maxmin fair allocation maximizes the smallest rate of any feasible rate allocation; given the best smallest rate allocation, it maximizes the next smallest allocation, etc.

The rate allocation obtained in such a way is fair in the sense that all flows constrained by a particular bottleneck get an equal share of this bottleneck capacity. It is also efficient in the sense that given the fair allocation, no more data can be pushed through the network, since each flow crosses at least one fully saturated link, which is the "bottleneck" link for that flow.

## 1.4 Related Work and Summary of Results

The procedure for achieving maxmin optimal rates described earlier used global information, which is expensive and difficult to maintain in the real-world networks.

Several feedback schemes have been proposed to achieve the same goal in a distributed network. In essence, all these schemes maintain some link controls at the switch and convey some information about these controls to the source by means of feedback. Upon receipt of the feedback signal the source adjusts its estimate of the allowed transmission rate according to some rule.

These algorithms essentially differ in the particular choices of link controls and the type of feedback provided to the source by the network.

References [7], [17], [19] describe distributed algorithms of this type. However, these algorithms required synchronization, which is difficult to achieve.

Mosley in [31] suggested an asynchronous algorithm for distributed calculation of maxmin fair rates. However, the algorithm convergence time was rather long and simulations showed poor adaptation to dynamic changes in the network.

Later Ramakrishnan, Jain and Chiu in [35] suggested a Selective Binary Feedback scheme (SBF) - a modification of the DECbit scheme with a calculation of fair allocation by the switch. The congestion bit is set only for flows exceeding the fair allocation, thus ensuring that maxmin fairness is achieved. Being still restricted to one bit of feedback only, SBF still produces oscillations and converges slowly compared to the scheme described in this paper.

Although it is mentioned in [35] that replacing the bit by a rate field would improve the performace of the scheme, it should be noted that this replacement would not yet eliminate the necessity of additive increase policy at the source. Without the additive increase policy SBF would not ensure convergence to maxmin fair solution. It is the contribution of [3] to note that at least an additional control bit is needed to ensure convergence if additive increase is eliminated. Replacing the additive increase policy by explicit rate setting is desirable for faster convergence.

In addition, our scheme requires significantly fewer iterations per a single switch allocation calculation: at most 2 for our scheme compared to N for SBF, where N is the number of different rates of flows traversing a given link (which is in the worst case as large as the number of different flows).

Thus, in the context of previous work, our approach is characterized by explicit calculation of optimal rates, and communication of these rates to the source, no requirement for any synchronized actions within the network, more rapid convergence, and accounting for the bandwidth used by the feedback traffic.

## 2  Global Calculation of Optimal Flow Rates

This section presents an analytical development of the global maxmin computation, and extends the basic idea of global computation of maxmin rates found in [2], [31], [17], [35] to account for the bandwidth consumed by feedback traffic, under the assumption that the feedback rate is proportional to the data rate in the forward direction.

Suppose there are $f$ forward and $b$ feedback flows traversing a given link. Let $k$ be the ratio of feedback to forward data rates. Then under the assumption that none of these flows are constrained by a bottleneck elsewhere in the network, the fair share of the link capacity $C$, allocated to each flow in its forward direction if $\frac{C}{f+kb}$. By computing this value for each link, we can find the *bottleneck* link in the network.

**Definition 2.1** *Within a network with links $\mathcal{L}$, we define a link $l$ as a* bottleneck link *if* $\frac{C_l}{f_l+kb_l} = \min_{j \in \mathcal{L}} \frac{C_j}{f_j+kb_j}$

Note that for $k = 1$ (feedback rate is equal to forward data rate) or for $k = 0$ (feedback rate is negligible compared to forward data rate), this definition reduces to that of [2].

Maxmin fair rates can now be found as follows:

- find all bottleneck links of the network and set the transmission rates of all the flows crossing these links in either direction to $\frac{C_l}{n_l}$ and mark those flows. Here and in what follows $n_l = f_l + kb_l$.

- decrease capacities of all links by the total capacity consumed by the marked flows crossing these links on their forward or feedback paths

- consider a reduced network with all link capacities adjusted as above and with marked flows removed. Repeat the procedure until all flows are assigned their rates and marked.

[3] contains a formal description of this procedure and a proof that it in fact yields maxmin fair rates.

## 3  Distributed Algorithm Description

Section 2 described a global synchronized procedure for determining the optimal rates of a fixed set of flows using the global knowledge of the network. This section presents a control scheme achieving the same goal in a distributed asynchronous way. Control information used for congestion management can be contained either in the data packet header or in special control packets (or cells in the ATM environment). In what follows we assume that special control packets are used.

Each source maintains an estimate of its optimal rate. Initially, it uses its desired sending rate as an estimate (perhaps infinity), but it updates this estimate by the periodic sending of control packets. The control packet contains two fields. The first field is one bit long and is called the "underloading" bit, or the "u-bit". The second field is several bits long and is used to contain the next rate estimate for the source. This field in the packet is called the "stamped rate."

When the source sends a control packet, it puts its current rate estimate in the "stamped rate" field, and clears the "u-bit" [1]. Each switch monitors its traffic and calculates its available capacity per flow. This quantity is called the "advertised rate". When a control packet arrives, the switch compares the "stamped rate" with the "advertized rate". If the "stamped rate" is higher than or equal to the "advertised rate", the stamped rate is reduced to the "advertised rate" and the "u-bit" is set. If the stamped rate is less than the advertised rate, the switch does not change the fields of the control packet.

When the control packet reaches the destination, the "stamped rate" contains the minimum of the source's rate estimate (at the time the control packet was sent) and all rates that the flow is allowed to have by the switches in its route. The destination sends the control packet back to the source. After a full round trip, the setting of the "u-bit" indicates whether the flow is constrained along the path. That is, if the "u-bit" is set, the rate is limited by some switch in the path and cannot be increased. In this case the source adjusts the "stamped rate" of its outgoing control packets to the "stamped rate" of the feedback control packet. If the "u-bit" is clear, the source's "stamped rate" is increased to its desired value.

The description of the computation of the "advertized rate" follows. The switches maintain a list of all of its flows and their last seen stamped rates, referred to as "recorded rates". The set of all flows whose "recorded rate" is higher than the switch's advertised rate are considered "unrestricted" flows and are denoted by $\mathcal{U}$. Similarly, flows with stamped rate below the advertised rate or equal to it are called "restricted flows" and are denoted by $\mathcal{R}$. The flows in $\mathcal{R}$ are assumed bottlenecked at some other switch or at this switch. Flows in the "unrestricted" set $\mathcal{U}$ are those for which a restricted rate has not yet been computed at this switch. Each switch, on receiving a control packet from a flow which is currently unrestricted at this switch, will compute a new "stamped rate" for this flow, under the assumption that this switch is a bottleneck for this flow. This will cause a switch to recompute its "advertized rate" as described below, and to insert a new rate into the "stamped rate" field of the control packet.

Given sets $\mathcal{R}$ and $\mathcal{U}$, the "advertized rate" is calculated as the link capacity not used by flows in $\mathcal{R}$

---

[1] If demand of a flow is below the allocation it receives from the network in the feedback packet, the source sets the "u-bit" of the outgoing control packets to 1.

available per per flow in $\mathcal{U}$. It is shown in [3] that there is some flexibility in exactly how the "advertized rate" is calculated. Here we give one possible way of doing it. Denote the "advertised rate" by $\mu$. Then it can be calculated as follows:

$$\mu = \frac{C - C_\mathcal{R}}{n - n_\mathcal{R}} \qquad (1)$$

where $C$ is the total capacity of the link, $C_\mathcal{R}$ is the total capacity consumed by all "restricted flows, $n = f + kb$, and $n_\mathcal{R} = f_\mathcal{R} + kb_\mathcal{R}$, with $f$, $b$, $f_\mathcal{R}$, $b_\mathcal{R}$ being the number of total and "restricted" forward and feedback flows traversing the link respectively. For $k = 1$ or $k = 0$, $n$ and $n_\mathcal{R}$ are simply the total number of flows and the number of "restricted" flows traversing the link.

It turns out that after this first recalculation of $\mu$ some flows that were previously "restricted" with respect to the old "advertised rate", can become "unrestricted" with respect to the new advertized rate. In this case these flows are re-marked as unrestricted and the advertised rate is recalculated once more. It is shown in [3] that the second recalculation is sufficient to ensure that any flow marked as restricted before the second recalculation remains restricted with respect to the newly calculated advertised rate.

Note that the value of the "stamped rate" the source writes in the control packet does not have to reflect the actual transmission rate at all times. Section 4 contains a discussion of different policies a source can implement to adjust its actual data transmission rate in response to the "stamped rate" received in feedback. In particular, if the desired value is unknown or very large, the actual transmission rate should not be increased when the "u-bit" is clear, while the "stamped rate" is set to the large value. When all flows stabilize to their optimal rates the "u-bit" will always be set when a contol packet returns to the source.

The formal description of the scheme can be found in [3].

### 3.1 M-consistency

The two-step calculation of the "advertized rate" in the previous section gave a particular example of how it might be calculated. Note that the result of this calculation is not only the advertized rate but also the sets $\mathcal{U}$ and $\mathcal{R}$ of "unrestricted" and "restricted" flows.

It turns out that there is some flexibility in choosing a particular policy for this calculation which will ensure convergence of the scheme to maxmin optimal rates.

We say that a calculation policy of the advertised rate is "M-consistent" (for "marking consistency") if after any update of the state of the switch the following conditions hold for any of its outgoing links:

1. if at any time a flow $j$ is marked to be in the "restricted set" $\mathcal{R}$, then its recorded rate $\xi$ does not exceed the "advertised rate" $\mu$

2. Given set $\mathcal{R}$, the "advertised rate" $\mu$ satisfies condition (1).

Essentially, M-consistency means that once the advertized rate is calculated, no flows remain marked as "restricted" with recorded rate exceeding the advertized rate.

It is shown in [3] that the two-step calculation of the previous section is M-consistent. The calculation of the "fair allocation" of the Selective Binary Feedback Scheme can also shown to be M-consistent.

The following section argues that the algorithm described in the previous section converges to maxmin optimal rates with any M-consistent calculation of the "advertised rate". We emphasize that preserving M-consistency is vital for ensuring the convergence and robustness properties of the scheme.

## 4 Convergence Properties

**Theorem 4.1** *Given arbitrary initial conditions on the states of all links in the network, states of all sources, destinations and arbitrary number of packets in transit with arbitrary control information written on them, the algorithm given in section 3 with any M-consistent calculation of the "advertized rate" converges to the optimal rates as long as the set of flows, their demands and routes eventually stabilize.*

The formal proof of this theorem is given in [3]. Here we will attempt to give an informal argument to provide the intuition on why it holds. For simplicity we assume that the demands of all flows are infinite. Note the case of finite demand can be reduced to the case of infinite demand by introducing artificial links at the source of the capacity equal to the demand.

Let $t_0$ denote the time by which all flows have become known at all links in their routes. It is shown in [3] that M-consistency of the advertised rate calculation implies that for all links $l$ for all times $t \geq t_0$

$$\mu_l \geq \frac{C_l}{n_l} \quad (2)$$

where $n_l = f_l + kb_l$.

This equation essentially means that the advertised rate is at least as large as the maxmin fair share per flow of this link would have been if this link were the tightest bottleneck in the network.

Let $\mathcal{L}_i$, denote the set of bottleneck links of the reduced network of iteration $i$, $\mathcal{S}_i$ denote flows crossing $\mathcal{L}_i$, and $\tau_i$ denote the optimal rates of theses flows. Let $\hat{\mathcal{L}}_i$ denote the set of links in $\mathcal{L} \setminus (\mathcal{L}_1 \cup \ldots \cup \mathcal{L}_i)$ s.t. at least one flow of $\mathcal{S} \setminus (\mathcal{S}_1 \cup \ldots \cup \mathcal{S}_i)$ crosses $l$.

It is shown in [3] that the following properties hold:

**Property 4.1** $\tau_1 < \ldots < \tau_m$

**Property 4.2** *Any flow in $\mathcal{S}_i$ traverses at least one link in $\mathcal{L}_i$ and only flows from $\mathcal{S}_1 \cup \ldots \cup \mathcal{S}_i$ traverse any link in $\mathcal{L}_i$ $\forall 1 \leq i \leq m$*

**Property 4.3** *Let $\hat{\mathcal{L}}_i$ denote the set of links $l \in \mathcal{L} \setminus \mathcal{L}_1 \cup \ldots \cup \mathcal{L}_i$ s.t. at least one flow of $\mathcal{S} \setminus \mathcal{S}_1 \cup \ldots \cup \mathcal{S}_i$ traverses $l$. Denote $n_l^j = f_l^j + kb_l^j$, where $f_l^j$ and $b_l^j$ are the number of flows of $\mathcal{S}_j$ crossing link $l$ in the forward and feedback direction respectively. Then $\forall 1 \leq i \leq m$*

$$\tau_i \begin{cases} = \frac{C_l - \sum_{j=1}^{i-1} \tau_j n_l^j}{n_l - \sum_{j=1}^{i-1} n_l^j} & \text{if } l \in \mathcal{L}_i \\ < \frac{C_l - \sum_{j=1}^{i-1} \tau_j n_l^j}{n_l - \sum_{j=1}^{i-1} n_l^j} & \text{if } l \in \hat{\mathcal{L}}_i \end{cases}$$

It can be easily seen that (2) and Properties 4.1-4.3 imply that

$$\mu_l(t) > \tau_1 \quad \forall l \in \mathcal{L} \setminus \mathcal{L}_1 \quad (3)$$
$$\mu_l(t) \geq \tau_1 \quad \forall l \in \mathcal{L}_1 \quad (4)$$

Consider now any flow $i \in \mathcal{S}_1$. By Property 4.2 it must traverse at least one of the bottleneck links $l \in \mathcal{L}_1$. Let $\mu_l(t_0)$ be the advertised rate of link $l$ at time $t_0$. Consider any control packet of $i$ which was sent on or after time $t_0$ with some stamped rate $\rho^0$. By operation of the algorithm when this packet returns to the source, two cases are possible:

Case 1. $\rho^0$ was greater than or equal to the advertized rate of at least one of the links in its route. Then, the stamped rate of the control packet will be reset to some $\rho^1$ which is the smallest advertised rate seen by the packet in its route, and its "u-bit" will be set to 1. Therefore, (3) and (4) imply that upon return the source $\rho_1 \geq \tau_1$, and therefore, after the first control packet round trip after time $t_0$ the stamped rate of all outgoing control packets will be greater than or equal to $\tau_1$.

Case 2. $\rho_0$ was smaller than the advertised rate of all links in its route. Then, the control packet will return with its "u-bit" set to 0. The next control packet sent out after that will have its stamped rate set to the demand of this flow (which is infinity). Repeating now the argument of Case 1 it should be clear that after the second round trip the stamped rate of all outgoing control packets will be greater than or equal to $\tau_1$.

We have shown that at most two control packet round trips after all flows become known at all links the stamped rate $\rho_i$ of any control packet of any flow $i$ satisfies

$$\rho_i \geq \tau_1 \quad \forall\ i \in \mathcal{S}_1 \qquad (5)$$
$$\rho_i > \tau_1 \quad \forall i \in \mathcal{S} \setminus \mathcal{S}_1 \qquad (6)$$

Therefore, after at most 3 roundtrips any switch l in the network will have recorded rates $\xi_i^l$ for all flows $i$ crossing this switch satisfying the condition:

$$\xi_i^l \geq \tau_1 \quad \forall\ l\ in\ the\ route\ of\ i \in \mathcal{S}_1 \qquad (7)$$
$$\xi_i^l > \tau_1 \quad \forall\ l\ in\ the\ route\ of;\ i \in \mathcal{S} \setminus \mathcal{S}_1 \qquad (8)$$

We will now show that after at most 3 roundrips

$$\mu_l = \tau_1\ \forall\ l \in \mathcal{L}_1 \qquad (9)$$

To see this consider any link $l \in \mathcal{L}_1$. By Property 4.2 only flows from $\mathcal{S}_1$ cross any link of $\mathcal{L}_1$.

Consider first the case where not all flows known at $l$ are in the "restricted" set $\mathcal{R}$. Then

$$\mu_l = \frac{C_l - \sum_{j \in \mathcal{R}_l} \xi_j^l \delta_j}{n_l - \sum_{j \in \mathcal{R}_l} \delta_j}$$

Here and in what follows $\delta_j = 1$ if $j$ is a forward flow and $\delta_j = k$ if $j$ is a feedbackflow.

Since $\tau_1 = \frac{C_l}{n_l}$ for any $l \in \mathcal{L}_1$,

$$\mu_l = \frac{\tau_1 n_l - \sum_{j \in \mathcal{R}_l} \xi_j^l \delta_j}{n_l - \sum_{j \in \mathcal{R}_l} \delta_j} \qquad (10)$$

holds for any $l \in \mathcal{L}_1$.
Note that (10) and (7), (8) imply that $\forall\ l \in \mathcal{L}_1$

$$\mu_l = \frac{\tau_1 n_l - \sum_{j \in \mathcal{R}_l} \xi_j^l \delta_j}{n_l - \sum_{j \in \mathcal{R}_l} \delta_j} \leq$$
$$\leq \frac{\tau_1 n_l - \tau_1 \sum_{j \in \mathcal{R}_l} \delta_j}{n_l - \sum_{j \in \mathcal{R}_l} \delta_j} = \tau_1$$

By (2) $\mu_l \geq \tau_1$ for all $l \in \mathcal{L}_1$, so (9) follows.

The case when all flows are in $\mathcal{R}$ is quite similar. Finally, it can easily be seen now that when all flows receive their third feedback control packet stamped rates of any control packet of any flow in $\mathcal{S}_1$ will be set to $\tau_1$. In turn, that implies that at most one control packet round-trip later all flows in $\mathcal{S}_1$ will be marked as "restricted" and their "recorded rates" $\xi$ will be set to $\tau_1$ at any switch along their route. Moreover, the advertized rate of any link will not be below $\tau_1$.

That is, we have shown that after at most 4 roundrtips all flows in $\mathcal{S}_1$ have reached their optimal rates *and these rates are no longer changed* provided the network conditions have not changed.

This fact permits us to consider the "reduced" network in which the flows of $\mathcal{S}_1$ are removed and the capacities of any link is decreased by $\tau_1$ multiplied by the number of flows of $\mathcal{S}_1$ crossing this link.

Now all of the arguments above can be repeated and thus we can show by induction that eventually all flows will be assigned their optimal rates.

Note that the proof of the convergence theorem outlined here can be used to obtain the upper bound on convergence time. Suppose that the round-trip delay of control packets is bounded by some $D$. Note that it takes at most 4D for all flows to complete each iteration and that there exactly as many iterations as there are bottlenecks of different capacity per flow restricted by that bottleneck (that is, there are as many iterations as the number of *different* rates in the optimal rate vector). Thus, the following Proposition holds:

**Proposition 4.1** *Given an upper bound $D$ on round-trip delay and the number $N$ of iterations of the global procedure, the upper bound on the algorithm convergence time is given by $4ND$.*

It can be easily seen that N is the number of different bottleneck rates. Note that $D$ includes the interval between sending control packets.

This bound gives the theoretical worst-case guarantee. In practice, convergence time should be expected to be significantly better. In fact, in our simulations we were not able to produce convergence worse than 2ND.

It is essential to note that the convergence time measured in round-trips of control packets does not give a good insight into the actual convergence measured in real time units if the time of round-trip delay $D$ is not satisfactory bounded. Given a feasible set of transmission rates, a network configuration, a particular underlying service discipline, and the source's traffic shaping mechanism, we could hope to be able to obtain such bound either from experiment or from theoretical analysis. References [33], [13] provide such upper bounds for particular service disciplines and source traffic shapes.

### 4.1 Setting the Actual Sending Rate

Unless special measures are taken, the transient infeasibility of transmission rates can cause significant queue growth and potential data loss. If control packets are of the same priority as data traffic, then transient infeasibility can drastically increase the upper bound on $D$ and cause slow convergence of the algorithm (as measured in real time rather than in the number of round-trips of control packets). Thus it would be very important to ensure that the algorithm produces a feasible transmission vector as early as possible.

Suppose that the network is started with some set of feasible actual transmission rates. Then we believe that the policy described below will preserve the feasibility of transmission rates. The key point is that the *actual* transmission rate does not have to be adjusted at the same time as the stamped rate. Suppose the value of $D$ for the network with *feasible* transmission rates is available for all sources. Then,

- if the "stamped" rate of the feedback control packet received at the source is below the current actual transmission rate, and the "u-bit" of the packet is set, then decrease the actual rate to the value of the "stamped" rate;

- if the "stamped" rate is greater than the current actual transmission rate and the "u-bit" is set, wait for $2D$ before increasing the actual transmission rate

- if the "u-bit" is not set, do not change the actual transmission rate.

The rationale for this policy is that decreasing the rate cannot possibly violate feasibility, while increasing it can, if the other flows are not yet aware of the increase. Since the *stamped* as opposed to the *actual* rate is in fact increased according to the original algorithm, on the next round-trip the new rate increase will be known at all links, so the other flows will be notified about this change no later than on their next round-trip after that. It should be also noted that two round-trips after all flows become known at all links, any flow's stamped rate is at least as large as the minimum of its demand and the equal share of the capacity of the bottleneck flow for this link. Note that while this may be less than the optimal rate of this flow, this bound ensures that all flows are guaranteed reasonable throughput even before the optimal rates are obtained.

The above considerations in combination with the simulation results presented in the next section lead us to believe that the algorithm is "well-behaved" in transience.

Finally it should be noted that the scheme proposed in this paper converges significantly faster than related work [35] and [31]. This becomes intuitively clear by noting that [35], while taking information about individual flows into account, does not have a way to efficiently use this information, since the source is only notified whether its rate must be increased or decreased, but is not told "how far to go". The scheme of [31] does inform the source about the rate it should transmit at, but the information needed to compute this rate is based on the aggregate and the maximum rates of all flows only. In contrast, our algorithm takes full information about the individual flow rates into account when calculating the rate estimate, and informs the source about this rate.

## 5 Simulation Results

The NETSIM simulation package developed at MIT was used in this work. [3] describes a number of simulation experiments performed on the number of different configurations. Here we present only one, which is to our mind the most representative.

Configuration of this experiment is shown in Figure 1. We investigate the behavior of the scheme in the case when flows enter and exit and when different flows have different number of links in their routes. In addition, there are 3 levels of bottleneck links here.

Optimal rates of flows at different times are given in Table 1. Figure 2 shows that all flows quickly determine their optimal rates after load changes.

## 6 Conclusions

We have described an algorithm for traffic management in which rate allocation is performed by the switches

| Flow/time | 0-15 | 15-48 | 48-67 | 67-100 |
|-----------|------|-------|-------|--------|
| 1 | 40 | 30 | - | 50 |
| 2 | 20 | 30 | - | - |
| 3 | 20 | - | - | - |
| 4 | 20 | 30 | 60 | 50 |
| 5 | 60 | 60 | 60 | 50 |

Table 1: Optimal rates of flows 1-5 at different times.

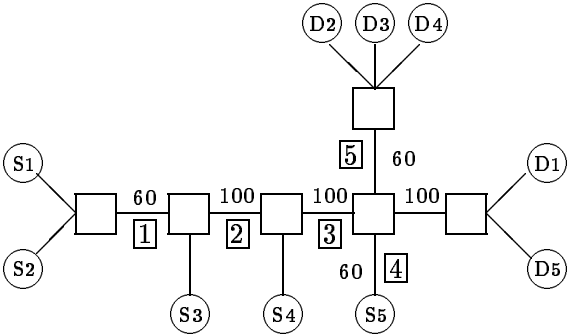

Figure 1: Configuration

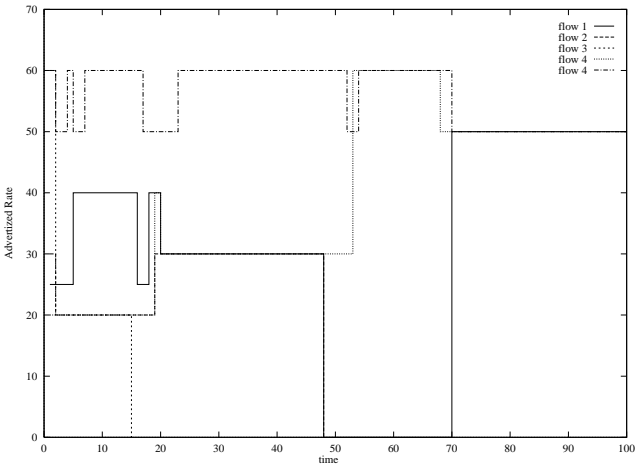

Figure 2: "Advertized" rate calculated by the algorithm. All 5 flows start at time 0. Then flow 3 exits at time 15. At time 48 flows 1 and 2 exit, and, finally, at time 67 flow 3 reenters. Demands of all flows are 70.

and the calculated rates are explicitly communicated to the sources in packet headers. While introducing obvious overhead, this approach has enabled us to achieve several important benefits; most notably, we have

- ensured fast convergence and robustness of the scheme

- eliminated frequent rate oscillations seen in schemes like DECbit

- reduced the computational complexity of the switch calculation compared to Selective Binary Feedback scheme

- eliminated the necessity of rate measurements at the switch

- made policing at the entry to the network more tractable, since the entry switch knows the value of the allowed rates for all flows by observing the "stamped rate" of the returning control packet

- allowed significant flexibility in the underlying service discipline

- Decoupled the rate calculation algorithm from the underlying network policies, thus allowing the algorithm to be run on top of other congestion control algorithms, in order to provide guidance to them on the transmission rates

It was suggested in [3] that the algorithm can be easily extended to the case where instead of end-to-end feedback the switch sends a feedback packet to the upstream switch to inform it about its control value if it is smaller. Such shortcuts will propagate the minimum link control value (advertized rate) to the source faster, provided the bottleneck is closer than at the very end of the route.

Another possibile way to improve the convergence time of the algorithm might be to restrict the allowed transmission rates to some discrete values. While this would certainly bound the number of potentially different values in the optimal vector and thus improve convergence time, the effects of such discretization need further research.

We believe that additional research is needed to investigate the transient behavior of the algorithm with different underlying network and higher level application policies.

# Acknowledgements

We would like to express our gratitude for the generous help of Jon C. R. Bennett in the preparation of this paper. We are grateful to K.K. Ramakrishnan for useful discussions of the Selective Binary Feedback scheme. We would also like to thank William Hawe for his help and useful comments.
# References

[1] "ISDN - Core Aspects of Frame Protocol for Use with Frame Relay Bearer Service", ANSI T1.618, 1991.

[2] Bertsekas, D., Galager, R., 1992. Data Networks, ch. 6, Prentice Hall, Englwood Cliffs, N.J.

[3] Charny, A., "An Algorithm for Rate Allocation in a Packet-Switching Network with Feedback". MIT/LCS/TR-601. April 1994.

[4] Clark, D., Schenker, S., Zhang, L. "Supporting Real-Time Applications in an Integrated Services Packet Network: Architecture and Mechanism", Proc. ACM SIGCOMM '92, Baltimore, August 1992, pp 14-26.

[5] Clark, D., Tennenhouse, D., " Architectural Considerations for a New Generation of Protocols, ", Proc ACM SIGCOMM '90, Vol. 20, No. 4, Philadelphia, September 1990, pp 200-208.

[6] Demers, A., Keshav, S., Shenker, S. 1990. "Analysis and Simulation of a Fair Queuing Algorithm", Internetwork: Research and Experience, Vol.1, No. 1, John Wiley & Sons, pp 3-26.

[7] Gafni, E., 1982. The Integration of Routing and Flow Control for Voice and Data in Integrated Packet Networks", Ph.D. thesis, MIT, Dept of Electrical Engineering and Computer Science, Cambridge, MA.

[8] Gafni, E., and Bertzekas, D., 1984 "Dynamic Control of Session Input Rates in Communication Networks", IEEE Trans. Automat. Control, AC-29, pp. 1009-1016.

[9] Gerla, M., Chan, W., de Marca, J.R.B "Fairness in Computer Networks", Proc. IEEE Int. Conf. Commun., June 1985, pp.1384-1389.

[10] Gerla, M., Kleinrock, L. 1980, "Flow Control: A Comparative Survey," IEEE Trans. Com., COM-28, pp.553-574.

[11] Golestani, S. 1980. "A Unified Theory of Flow Control and Routing on Data Communication Networks", Ph.D. thesis, MIT, Dept of Electrical Engineering and Computer Science, Cambridge, MA.

[12] Golestaani, S. "Congestion Free Real-Time Traffic in Packet Networks", Proc. of IEEE INFOCOM '90, San Francisco, Cal., June 1990, pp. 527-536.

[13] Golestani, S. "A Stop-and-Go Queuing Framework for Congestion Management", Proc. of ACM SIGCOM '90, Vol. 20, No.4, September 1990, pp.8-18.

[14] Hahne, E., Gallager, R., "Round-Robin Scheduling for Fair Flow Control in Data Communication Networks", (report LIDS-P-1537). Cambridge, MA: MIT Laboratory for Information and Decisions Systems.

[15] Hahne, E., 1986. "Round-Robin Scheduling for Fair Flow Control in Data Communication Networks", Ph.D. thesis, MIT, Dept of Electrical Engineering and Computer Science, Cambridge, MA.

[16] Hahne, E."Round-Robin Scheduling for Maxmin Fairness in Data Networks", IEEE J. on Sel. Areas in Comm., vol 9, No.7, September 1991.

[17] Hayden, H. 1981. "Voice Flow Control in Integrated Packet Networks', (report LIDS-TH-1152). Cambridge, MA: MIT Laboratory for Information and Decisions Systems.

[18] Jaffe, J. 1981. "A Decentralized Optimal Multiple-User Flow Control Algorithm", Proc. Fifth Int. Conf. Comp. Comm, October 1980, pp 839-844

[19] Jaffe, J. "Bottleneck Flow Control", IEEE Trans. Comm., COM-29, No 7. pp954-962.

[20] Jacobson, V. "Congestion Avoidance and Control" , Proc. ACM SIGCOMM '88, Stanford, Calif., August 1988, pp.314-329.

[21] Jain, R., Chiu D.-M., Hawe, W., " A Quantitative Measure of Fairness and Discrimination for Resource Allocation in Shared Computer System", DEC-TR-301, Digital Equipment Corporation. 1984

[22] Jain, R., Ramakrishnan, K., " Congestion Avoidance in Computer Networks With a Connectionless Network Layer. Part 1: Concepts, Goals and Methodology", DEC-TR-507, Digital Equipment Corporation. 1987

[23] Jain, R., "Myths about Congestion Management in High-Speed Network," DEC-TR-724, Digital Equipment Corporation. 1990.

[24] Nagle, J., "On Packet Switches with Infinite Storage", IEEE Trans. on Comm., Vol.35, No.4, April 1987, pp. 435-438

[25] Kanakia, H., Mishra, P., Reibman, A., "An Adaptive Congestion Control Scheme for Real-Time Packet Video Transport", SIGCOMM'93, October 1993.

[26] Keshav, S. "A Control-Theoretic Approach to Flow Control", SIGCOMM'91, September 1991.

[27] Keshav, S., Agrawala, A., Singh,S, "Design and Analysis of A Flow Control Algorithm for a Network of Rate Allocating Servers", International Workshop on Protocols for High-Speed Networks, Palo Alto, Cal., November, 1990.

[28] Kleinrock, L., "Queuing Systems", vol.1., New York, Wiley, 1975.

[29] Kleinrock, L., "Queuing Systems", vol.2., New York, Wiley, 1976.

[30] Mishra, P., Kanakia, H.,"A Hop by Hop Rate-based Congestion Control Scheme", SIGCOMM'92, pp. 112-123, October 1992

[31] Mosley, J. 1984. " Asynchronous Distributed Flow Control Algorithms", Ph.D. thesis, MIT, Dept of Electrical Engineering and Computer Science, Cambridge, MA.

[32] Partridge, C., 1993. Gigabit Networking, ch. 3,4,7,8,11-13, Addison-Wesley Publishing Co, Inc.

[33] Parekh, A., 1992. " A Generalized Processor Sharing Approach to Flow Control in Integrated Services Networks," Ph.D. thesis, MIT, Dept of Electrical Engineering and Computer Science, Cambridge, MA.

[34] K.K. Ramakrishnan, Raj Jain. Congestion Avoidance in Computer Networks with a Connectionless Network Layer. Part II. An Explicit Binary Feedback Scheme. DEC-TR-508, 1987

[35] K.K.Ramakrishnan, Raj Jain, Dah-Ming Chiu. " Congestion Avoidance in Computer Networks With a Connectionless Network Layer. Part IV: A Selective Binary Feedback Scheme for General Topologies Methodology", DEC-TR-510, Digital Equipment Corporation. 1987

[36] Rose, O., "The Q-bit scheme: congestion avoidance using rate-adaption", Computer Communication Review vol.22, no.2 p.29-42, April, 1992.

[37] Shenker, Scott, "A Theoretical Analysis of Feedback Flow Control", Proc. SIGCOMM'90, Philadelphia, PA, September 1990, pp.156-165.

[38] Zhang, L., Shenker, S., Clark, D., " Observations on the Dynamics of a Congestion Control Algorithm: The Effects of Two-Way Traffic," Proc. ACM SIGCOMM '91, Zurich, September 1991, pp.133-148.

[39] Yin,N.,Hluchyj,M. "A Dynamic Rate Control Mechanism for Source Coded Traffic in a Fast Packet Network", IEEE Journal on Selected Areas in Communications, Vol.9, No.7, pp.1003-1012, Sept.1991